\documentclass[preprint,1p,12pt,sort&compress]{elsarticle}
\usepackage{graphicx}
\usepackage{amsmath,amssymb}
\usepackage[dvipdfm, colorlinks=true, pdfstartview=FitV, linkcolor=red, citecolor=blue, urlcolor=blue]{hyperref}
\usepackage{color}

\newcommand{\vect}[1]{\mathbf{#1}}

\newcommand{\im}{\mathrm{Im}\,}
\newcommand{\re}{\mathrm{Re}\,}
\newcommand{\sgn}{\mathrm{sgn}}
\newcommand{\Slash}[1]{\ooalign{\hfil/\hfil\crcr$#1$}}
\newcommand{\vzero}{\vect{0}}
\newcommand{\vp}{\vect{p}}
\newcommand{\vk}{\vect{k}}
\newcommand{\vgamma}{{\boldsymbol \gamma}}

\newcommand{\GR}{G^R}
\newcommand{\DR}{D^{R}}

\newcommand{\GF}{G^{S}}
\newcommand{\DF}{D^{S}}

\newcommand{\mb}{m_b}
\newcommand{\mf}{m_f}

\newcommand{\cp}{g}
\newcommand{\kernel}{K}
\newcommand{\selfEnergyR}{\Sigma^R}

\newcommand{\deltam}{\delta m}

\newcommand{\mph}{m_\gamma}
\newcommand{\me}{m_e}
\newcommand{\zetae}{\zeta_e}
\newcommand{\zetaph}{\zeta_\gamma}
\newcommand{\damping}{\zeta}
\newcommand{\pzero}{p^0}
\newcommand{\pw}{(\pzero,\vp)}

\usepackage{ulem}

\newcounter{nombre}

\begin{document}
\vspace*{-30mm}
\begin{flushright}
{\scriptsize RIKEN-MP-37}
\end{flushright}
\vspace{5mm}
\begin{frontmatter}
\title{Ultrasoft Fermionic Modes  at High Temperature}

\author[riken]{Yoshimasa Hidaka}
\ead{hidaka@riken.jp}
\author[kyoto]{Daisuke Satow}
\ead{d-sato@ruby.scphys.kyoto-u.ac.jp}
\author[kyoto]{Teiji Kunihiro}
\ead{kunihiro@ruby.scphys.kyoto-u.ac.jp} 

\address[riken]{Mathematical Physics Laboratory, RIKEN Nishina Center, Saitama 351-0198, Japan}
\address[kyoto]{Department of Physics, Faculty of Science, Kyoto University, Kitashirakawa Oiwakecho, Sakyo-ku, Kyoto 606-8502, Japan}

\begin{abstract}
A possible collective fermionic excitation in the ultrasoft energy-momentum region $p \lesssim \cp^ 2T$
is examined in Yukawa model with  scalar coupling and quantum electrodynamics (QED) 
with $\cp$ being coupling constant 
at extremely high temperature $T$ where the fermion mass is negligible.
We  analytically sum up the ladder diagrams for the vertex correction in the leading order in QED,
 which is not necessary in the Yukawa model, and find that the fermion pole exists at  $\omega = \pm |\vp|/3-i\zeta$ 
with ultrasoft momentum $\vp$ both for the Yukawa model and QED; 
 $\zeta$ is the sum of the damping rates of fermion and boson with hard momenta.
We also obtain the expression of the residue of the pole, 
which is as small as of order $\cp^2$.
We show that the fermion propagator and the vertex function
satisfy the Ward-Takahashi identity in QED.
Thus we establish the existence of  an ultrasoft
fermionic mode at extremely high temperature, 
which was originally called phonino and was suggested 
in the context of supersymmetry and its breaking at finite $T$. We discuss the possible origin of 
such an ultrasoft fermionic mode without recourse to supersymmetry.
The case of QCD is briefly mentioned.
\end{abstract}
\end{frontmatter}

\section{Introduction}
\label{sec:introduction}
\begin{figure}
\begin{center}
\includegraphics[width=0.55\textwidth]{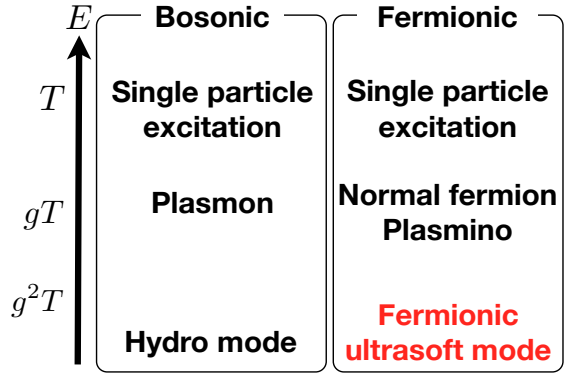}
\caption{Schematic illustration of the energy hierarchy of
bosonic and fermionic modes in a massless fermion-boson system at finite temperature:
The vertical axis is the energy scale.
The left  and right part display the bosonic and  fermionic modes, respectively.
The bosonic modes may be of scalar-, gauge-bosonic and hydrodynamic ones.
No established fermionic mode had been known in the ultrasoft region.  
}
\label{fig:collective}
\end{center}
\end{figure}

Revealing and clarifying the nature of 
possible quasiparticle and collective excitations 
is of  basic importance for understanding a many-body system, 
in particular, in the low-energy regime.
In fermion-boson systems such as  Yukawa model, quantum electrodynamics (QED), and quantum chromodynamics (QCD)
at so high temperature $T$ that the masses of the particles are negligible, 
the average inter-particle distance is proportional to $1/T$ while some collective effects can be expected 
in the soft momentum scale $1/\cp T$ with $\cp$ being the coupling constant.
Indeed, soft bosonic modes in the longitudinal as well as the transverse channels
exist and are known as plasmon~\cite{plasmon}, 
while the fermionic counter part is known as plasmino~\cite{HTL,plasmino}, 
both of which have masses of order $\cp T$.
In the lower energy region, there exist 
hydrodynamic modes of bosonic nature, which are actually the  zero modes 
associated with the conservation of energy-momentum and charges.
The energy hierarchy of bosonic and fermionic modes at high temperature 
may be summarized schematically as shown in Fig.~\ref{fig:collective}.
From the figure, one may have an intriguing but natural question whether such ultrasoft 
($\lesssim \cp^2T$) or zero modes can exist also in the fermion sector, 
possibly when the fermion system has a peculiar symmetry
 such as  chiral symmetry. 
In this paper, we argue and demonstrate that such a fermionic mode can exist in this infrared energy region
at extremely high temperature.

Here we should mention that there have been some suggestive works for supporting the existence of 
such an ultrasoft fermionic mode at finite temperature.
Historically, the ultrasoft fermionic  mode at finite $T$ was found in supersymmetric models 
as Nambu-Goldstone fermions called phonino associated with
spontaneously breaking of supersymmetry at $T\not=0$, 
which was shown by using Ward-Takahashi identity and a diagrammatic technique~\cite{phonino,lebedev1}. 
Here we note that the analysis in~\cite{phonino,lebedev1} 
was performed in the temperature region much below the critical temperature, 
$T_c\sim m/\cp$, where $m$ is the bare mass. It implies that their analysis is only valid for
$gT\ll m$, but not for $m\ll gT$ for which our analysis in the present paper
is concerned.
The analysis was extended to QCD at so high temperature that the coupling constant is weak~\cite{lebedev2}, 
in which a supersymmetry is still assigned at  the vanishing coupling, and hence, 
the supersymmetry is, needless to say, explicitly broken by the interaction.
Thus, there exists no exact fermionic zero mode  but only a pseudo-phonino does.
Although these analyses~\cite{phonino,lebedev1,lebedev2} are suggestive, 
it is still obscure whether a genuine ultrasoft fermionic mode exists
 when supersymmetry is absent, in particular, at extremely high temperature.

Here we note that there have been suggestions of the existence of ultrasoft
fermionic mode at finite $T$ even without supersymmetry.
It was shown in one-loop calculations \cite{3peak,mitsutani} 
that when a fermion is coupled with a massive boson with  mass $m$, 
the spectral function of the fermion gets to have  a novel peak in the far-low-energy region  in addition 
to the normal fermion and the plasmino, when $T\sim m$,
 irrespective of the type of boson; it means that 
 the spectral function of the fermion has a three-peak structure in this temperature region. 
Recently, the present authors \cite{shk} have suggested that such a three-peak
structure may persist even at the high temperature limit in the sense $m/T\rightarrow 0$, for the massive vector boson on the basis of a gauge-invariant formalism, again, at the one-loop order. 
Thus, one may expect that the novel excitation may exist in the far-infrared region 
also for a fermion coupled with a massless boson,
although the one-loop analysis admittedly may not be applicable at the ultrasoft momentum region.
There are also the works suggesting the existence of the ultrasoft fermionic mode using the Schwinger-Dyson 
equation~\cite{S-D}; we note, however,  that it is difficult to keep gauge symmetry in the
Schwinger-Dyson approach  at finite $T$.

Now let us give a generic argument supporting the existence of an ultrasoft fermionic
mode at finite $T$ on the basis of the symmetry of the self-energy for a massless fermion.
In this case, the important point is that the fermion is chiral;
the real part of the retarded fermion self-energy, $\Sigma^R(\omega,\vzero)$, 
at zero spatial momentum and the vanishing chemical potential,
is an odd function of $\pzero$:
\begin{align}
\re\Sigma^R(-\pzero,\vzero)=-\re\Sigma^R(\pzero,\vzero).
\end{align} 
If the $\re\Sigma^R(\pzero,\vzero)$ is  a smooth function at $\pzero=0$, then,
 $\re\Sigma^R(-\pzero,\vzero)=0$ at $\pzero=0$, 
which implies $\re G^{-1}(\pzero,\vzero)=0$ at $\pzero=0$;
the spectrum has a peak at the origin provided that the imaginary part of the fermion self-energy 
is not too large.
This argument suggests that the existence of the ultrasoft pole may be a universal phenomenon
 at high temperature in the theory composed of  massless fermion coupled with a boson.

It is, however,  not a simple task to establish that fermionic modes exist in the ultrasoft region on a general ground beyond the one-loop order accuracy 
because of the infrared divergence called pinch singularity 
\cite{lebedev2,pinchsingularity,transport,ultrasoft-am,hidaka} 
that breaks a naive perturbation theory, as will be briefly reviewed 
in the next section.
We remark that the same difficulty arises in the calculation of transport coefficients 
\cite{transport,hidaka} and the gluon self-energy \cite{ultrasoft-am} in the ultrasoft energy region.
Therefore, in this paper, we analyze the fermion propagator in the ultrasoft energy region in Yukawa model and QED
using a similar diagrammatic technique in Refs.~\cite{lebedev1,lebedev2,transport,hidaka} to regularize the pinch singularity.
We shall show that the retarded fermion propagator has a pole 
at $\pzero = \pm |\vp|/3-i\zeta$  ($\zeta\sim \cp^4T\ln \cp^{-1}$ for Yukawa model and $\sim \cp^2T\ln \cp^{-1} $ for 
QED)
with the residue $Z\sim \cp^2$ for ultrasoft momentum $\vp$ taking into account the ladder summation. 

This paper is organized as follows:
In Sec.~\ref{sec:resum}, we discuss the ultrasoft fermionic mode in Yukawa model as  a simple example without supersymmetry.
In Sec.~\ref{sec:QED}, we examine  the ultrasoft fermionic mode in QED.
We  analytically sum up the  ladder diagrams giving the vertex correction in the leading order,
 and find the existence of 
the ultrasoft fermionic mode as in the Yukawa model. We shall also show that the constructed propagator and the
vertex satisfy the Ward-Takahashi identity. 
Section~\ref{sec:discussion} is devoted to the discussion on the physical origin of the ultrasoft fermionic mode, and in Sec.~\ref{sec:summary}, we will summarize our results.

\begin{figure}
\begin{center}
\includegraphics[width=0.4\textwidth]{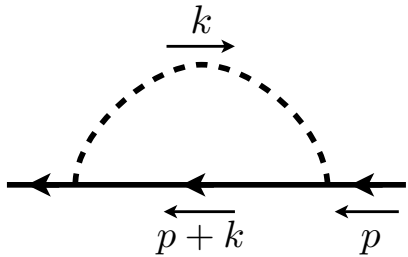}
\caption{Diagrammatic representation of the fermion self-energy in Eq.~(\ref{eq:one-loop-bare}) 
at one-loop order.
The solid and dashed line are the propagator of the fermion and the scalar  boson, respectively.
In Eq.~(\ref{eq:one-loop-pc}), the fermion and the boson propagator in the internal lines 
are replaced by the dressed propagators given in Eq.~(\ref{eq:resum-propagator}).}
\label{fig:oneloop}
\end{center}
\end{figure}

\section{Ultrasoft Fermionic Modes in Yukawa model}
\label{sec:resum}
Let us start with Yukawa model, which is the simplest model to study the ultrasoft fermionic modes.
Generalization to gauge theory will be discussed in Sec.~\ref{sec:QED}.
As stated in  Introduction,
we deal with a massless fermion (denoted by $\psi$) coupled with a
scalar field $\phi$ through the interaction Lagrangian 
$\mathcal{L}_I=g\phi\bar{\psi}\psi$.
We do not include the possible self-coupling of the scalar fields for simplicity.
We calculate the fermion retarded self-energy and
obtain the fermion retarded Green function 
with an ultrasoft momentum $p\lesssim \cp^2T$.
We first see that the naive perturbation theory breaks down
in this case. 
Then, we shall show that a use of a dressed propagator gives
a sensible result in the perturbation theory 
and that  the resulting fermion propagator has a new pole in the ultrasoft region.

\subsection{One-loop calculation}
The retarded self-energy in the one-loop level is given 
by
\begin{equation}
\label{eq:one-loop-bare}
\begin{split}
\selfEnergyR_{\text{bare}}(p)&= i\cp^2\int\frac{d^4k}{(2\pi)^4}\Bigl[\DF_0(-k)\GR_0(p+k)
+\DR_0(-k)\GF_0(p+k)\Bigr],
\end{split}
\end{equation}
where $D_0^{R,S}(-k)$ and $G_0^{R,S}(p+k)$ are the {\it bare} propagators of the fermion and the scalar boson defined as
\begin{align}
\label{eq:bare-propagatorS}
\GR_0(k)&=\frac{-\Slash{k}}{k^2+ik^0\epsilon},\\
\GF_0(k)&=\left(\frac{1}{2}-n_F(k^0)\right)i\Slash{k}(2\pi) \sgn(k^0)\delta(k^2),
\\
\DR_0(k)&=\frac{-1}{k^2+ik^0\epsilon},\\
\DF_0(k)&=\left(\frac{1}{2}+n_B(k^0)\right)i(2\pi)\sgn(k^0)\delta(k^2).
\label{eq:bare-propagatorE}
\end{align}
Here, $n_F(k^0)\equiv1/(\exp(k^0/T)+1)$ and $n_B(k^0)\equiv {1/(\exp(k^0/T)-1)}$ are 
the Fermi-Dirac and Bose-Einstein  distribution functions, respectively.
In the present analysis, we have preferentially employed
 the real-time formalism in Keldysh basis~\cite{Keldysh}. 
The diagrammatic representation of Eq.~(\ref{eq:one-loop-bare}) is shown in Fig.~\ref{fig:oneloop}. 
Inserting Eqs.~(\ref{eq:bare-propagatorS}) through (\ref{eq:bare-propagatorE})
into (\ref{eq:one-loop-bare}), we obtain
\begin{equation}
\begin{split}
\selfEnergyR_{\text{bare}}(p)&= \cp^2\int\frac{d^4k}{(2\pi)^4}\Bigl[\left(\frac{1}{2}+n_B(k^0)\right)\\
&\quad\times
\frac{\Slash{k}+\Slash{p}}{p^2+2p\cdot k+i(k^0+\pzero)\epsilon}(2\pi)\sgn(k^0)\delta(k^2)\\
&\quad-\left(\frac{1}{2}-n_F(k^0+\pzero)\right)
\frac{\Slash{k}+\Slash{p}}{p^2+2p\cdot k+ik^0\epsilon}\\
&\qquad\qquad\times
(2\pi)\sgn(k^0+\pzero)\delta((k+p)^2)\Bigr],
\end{split}
\end{equation}
where we have used the on-shell conditions for the bare particles, 
$k^2=0$, and $(k+p)^2=0$ in $\DF_0(-k)$ and  $\GF_0(p+k)$.
Then, for small $p$, the self-energy is reduced to
\begin{equation}
\begin{split}
\selfEnergyR_{\text{bare}}(p)&= \cp^2\int\frac{d^4k}{(2\pi)^4}\kernel(k)\frac{\Slash{k}}{2p\cdot k+ik^0\epsilon},
\label{eq:bareSelfEnergy}
\end{split}
\end{equation}
where 
\begin{equation}
\kernel(k)=(2\pi)\,\sgn(k^0)\delta(k^2)(n_F(k^0)+n_B(k^0)).
\end{equation}
Note that $\kernel(k)$ is independent of $p$.
This approximation is equivalent to the HTL approximation~\cite{HTL}.
The HTL approximation is, however, only valid for
$p\sim \cp T$, and not applicable in the ultrasoft momentum region.
In fact, the retarded self-energy in the one-loop level obtained with use of the bare
propagators is found to diverge when $p\rightarrow 0$,
since the integrand contains $1/ p\cdot k$.
This singularity is  called ``pinch singularity'' \cite{lebedev2,pinchsingularity,transport,ultrasoft-am,hidaka}.

The origin of this singularity is traced back to the use 
of the bare propagators because the singularity is caused by the fact that the dispersion 
relations of the fermion and the boson are the same and the damping rates are zero in these
 propagators.
For this reason,  one may suspect that this singularity can be removed
by adopting the dressed propagators taking into account the asymptotic masses and decay widths 
of the quasiparticles,
as will be shown to be the case shortly.

Since the leading contribution comes from the hard ($k \sim T$) internal and almost on-shell ($k^2\approx 0$)
momentum\footnote{Here we 
note that the case where the internal momenta are soft ($k\sim gT$) or smaller is not relevant:
In fact,   the HTL-resummed propagators \cite{HTL-resum} should be used for soft momenta.
However, the dispersion relations of the fermion and the boson obtained from these propagators 
are different from each other, so the pinch singularity will not appear in this case.},
we are led to employ the following dressed propagators for the fermion and boson:
\begin{align}
\label{eq:resum-propagator}
\GR(k)\simeq&-\frac{\Slash{k}}{k^2-m^2_f+2i\zeta_f k^0},\\
\GF(k)\simeq&\left(\frac{1}{2}-n_F(k^0)\right)\Slash{k}\frac{4i\zeta_fk^0}{(k^2-m^2_f)^2+4\zeta^2_f(k^0)^2},\\
\DR(k)\simeq&-\frac{1}{k^2-m^2_b+2i\zeta_b k^0},\\
\DF(k)\simeq&\left(\frac{1}{2}+n_B(k^0)\right)\frac{4i\zeta_bk^0}{(k^2-m^2_b)^2+4\zeta^2_b(k^0)^2} ,
\end{align}
where $\mf\equiv \cp T/(2\sqrt{2})$ and $\mb\equiv \cp T/\sqrt{6}$ are the  asymptotic masses  
of the fermion and the boson at $k^2\simeq0$, respectively~\cite{scalar,dispersion-numerical,Blaizot:2001nr}.
The damping rates of the hard particles, $\zeta_f$ and $\zeta_b$, are of order $\cp^4T\ln\cp^{-1}$. The logarithmic enhancement for the damping rate  is caused by the soft-fermion exchange,
which is analogous to that of the hard photon~\cite{damping-hard-photon}.
Note that these resummed propagators are the same as those used in \cite{lebedev2}, 
except for the smallness of the damping rates:
We remark that such a smallness is not the case in QED/QCD, where
the damping rate is anomalously large and of order $g^2T\ln g^{-1}$ (``anomalous damping'') \cite{damping-hard-electron}. 

Using these dressed propagators, we obtain
\begin{equation}
\label{eq:one-loop-pc}
\Sigma^R(p)\simeq \int\frac{d^4k}{(2\pi)^4}\tilde{\kernel}(k)
\frac{\Slash{k}}{1+2\tilde{p}\cdot k/\delta m^2 }
\end{equation}
for small $p$,
where $\delta m^2\equiv \mb^2-\mf^2=g^2T^2/24$, $\zeta\equiv \zeta_f+\zeta_b$, 
$\tilde{\kernel}(k)\equiv (g^2/\delta m^2) \kernel(k)$, and $\tilde{p}^\mu=(\pzero+i\zeta, \vect{p})$.
We have used the modified on-shell condition of  the quasi-particles,
$k^2-m_f^2+2i\zeta_f k^0=0$ and $k^2-m_b^2+2i\zeta_f b^0=0$, 
to obtain the denominator of the integrand in Eq.~(\ref{eq:one-loop-pc}).
We have also neglected $m_b$, $m_f$, $\zeta_b $, and $\zeta_f $ in $\kernel(k)$,
since the leading contribution comes from hard momenta $k\sim T$.
It is worth emphasizing that thanks to $\delta m^2$ and $\zeta$, 
$\Sigma^R(p)$ given in Eq.~(\ref{eq:one-loop-pc}) does not diverge in the infrared limit, $p\rightarrow 0$.

Before evaluating Eq.~(\ref{eq:one-loop-pc}),  we introduce the the following dimensionless value:
\begin{equation}
\lambda\equiv \int\frac{d^4k}{(2\pi)^4}\tilde{\kernel}(k) = \frac{\cp^2T^2}{8\deltam^2},
\end{equation}
which is of order unity. This value will characterize the strength of residue of the pole for both Yukawa model and QED.

We expand the self-energy in terms of $\tilde{p}^{\mu}$ instead of $p^{\mu}$ itself.
This is the key point of our expansion, which enable us to analytically find the pole of the ultrasoft fermionic mode.
Then,  the leading contribution is
\begin{equation}
\label{eq:sigma-p1}
\begin{split}
\Sigma^R(p)&\simeq-\int \frac{d^4k}{(2\pi)^4}\tilde{K}(k)\Slash{k}\frac{2\tilde{p}\cdot k}{\delta m^2}
=-\frac{1}{Z}\left((\pzero+i\zeta)\gamma^0+v \vp\cdot \vgamma\right),
\end{split}
\end{equation}
with $Z\equiv \cp^2/(8\lambda^2\pi^2)$ and $v=1/3$. 
Note that the zeroth-order term is absent,  which implies that there is no mass term.
Thus, we obtain the fermion propagator in the ultrasoft region as
\begin{equation}
\label{eq:result-propagator}
\begin{split}
\GR\pw=&-\frac{1}{\Slash{p}-\selfEnergyR\pw}
\simeq\frac{1}{\selfEnergyR\pw}\\
=&-\frac{Z}{2}\left(\frac{\gamma^0-\hat{\vp}\cdot\vgamma }{\pzero+v|\vp|+i\zeta}+\frac{\gamma^0+\hat{\vp}\cdot\vgamma}{\pzero-v|\vp|+i\zeta}\right).
\end{split}
\end{equation}
Here we have decomposed the fermion propagator into the fermion number $+1$ and $-1$ sectors
in the second line.
These two sectors are symmetric under $\vp\leftrightarrow -\vp$ and $v\leftrightarrow -v$,
so we analyze only the fermion number $+1$ sector in the following.

\begin{figure}[t]
\begin{center}
\includegraphics[width=0.4\textwidth]{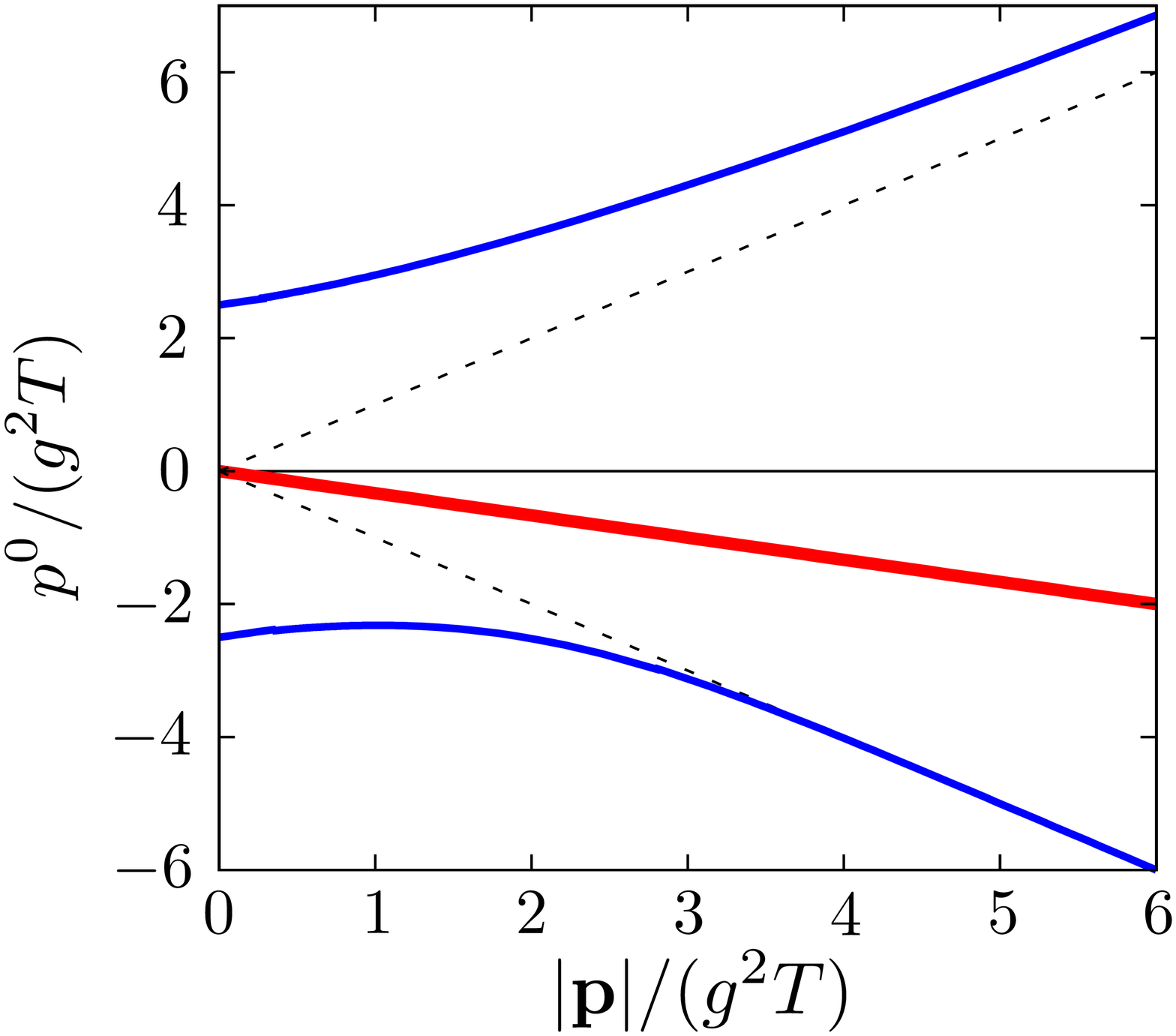}
\includegraphics[width=0.45\textwidth]{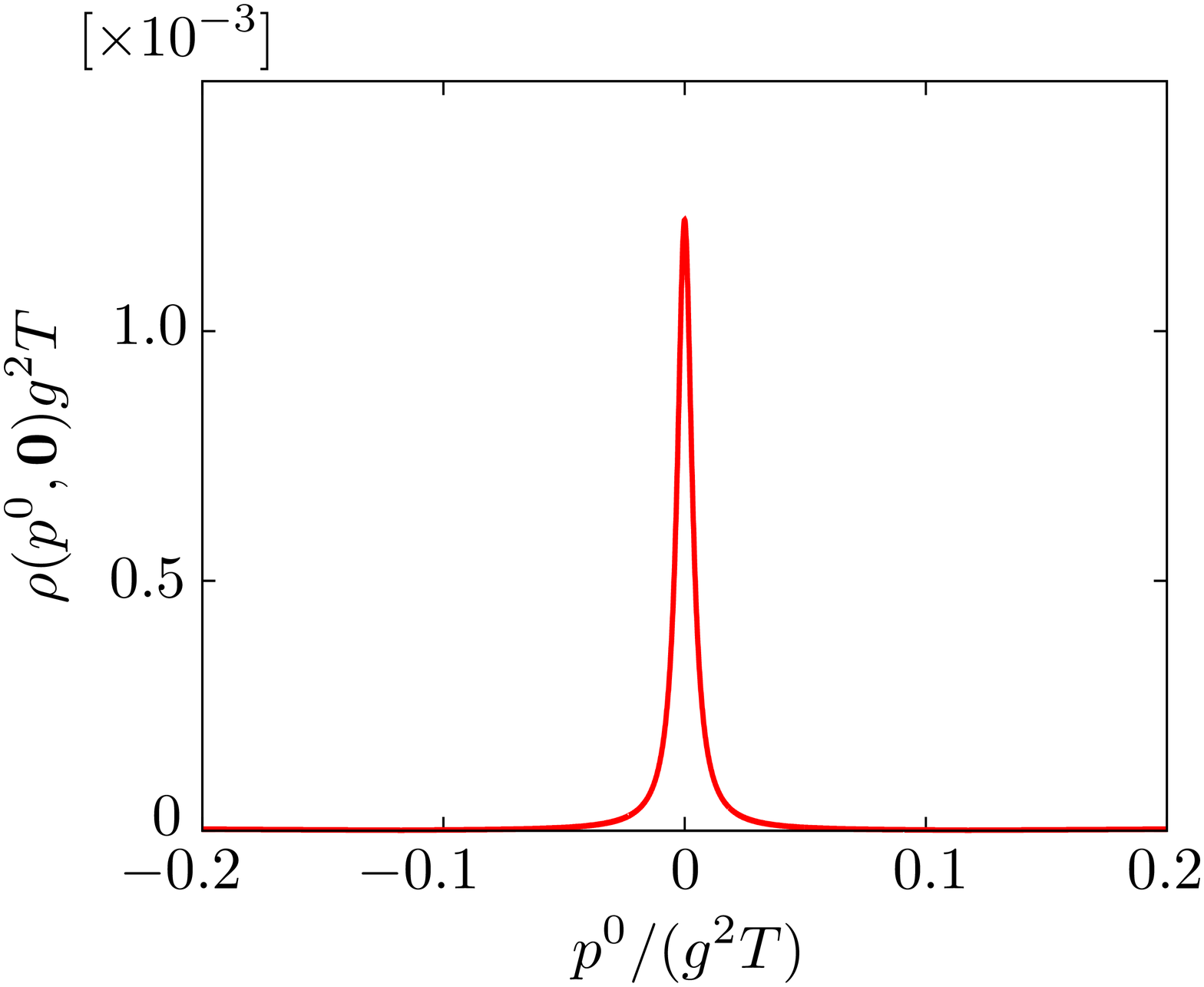}
\caption{Left panel: The dispersion relation in the fermionic sector.
In both of the two figures, the coupling constant is set to $\cp=0.1$.
The vertical axis is the energy $\pzero$, while the horizontal axis is the momentum $|\vp|$.
The solid (blue) lines correspond to the normal fermion and the antiplasmino, while the bold solid (red) one to the ultrasoft mode.
Note that since we focus on the fermion sector, the antiplasmino appears instead of the plasmino.
The dotted lines denote the light cone.
Since our analysis on the ultrasoft mode is valid only for $|\vp|\ll \cp^2T$, 
the plot for $|\vp|\gtrsim \cp^2T$ may not have a physical meaning.
The residue of the antiplasmino becomes exponentially small 
for $|\vp|\gg \cp T$, 
so the plot of the antiplasmino does not represent physical excitation for
 $|\vp|\gg \cp T$, either.
Right panel: The spectral function in the fermion sector, Eq.~(\ref{eq:spectral}), as a function of energy $\pzero$ at zero momentum.
}
\label{fig:dispersion}
\end{center}
\end{figure}

From Eq.~(\ref{eq:result-propagator}), we find a pole at 
\begin{align}
\label{eq:result-dispersion}
\pzero=- v|\vp|-i\zeta.
\end{align}
Note that the real part of $\pzero$ is negative for the fermion sector, which
 suggests that this peak has an antifermion-hole-like character
like the antiplasmino \cite{plasmino}.
The dispersion relation of the real part, $\re\pzero=-v|\vp|$, is shown 
in the left panel of Fig.~\ref{fig:dispersion} together with the HTL results \cite{HTL,plasmino} for comparison,
where the coupling constant is chosen as $\cp=0.1$.
The imaginary part of the pole reads
\begin{equation}
\label{eq:result-width}
\zeta \sim \cp^4T\ln g^{-1},
\end{equation}
which is much smaller than those of the normal fermion and the antiplasmino \cite{scalar}.
Since the real part and the imaginary part of the pole are finite for $|\vp|\neq 0$, 
this mode is a damped oscillation mode. 
The residue of the pole is evaluated to be
\begin{align}
\label{eq:result-residue}
Z =\frac{\cp^2}{8\lambda^2\pi^2}=\frac{\cp^2}{72\pi^2}\sim \cp^2,
\end{align}
which means that the mode has only a weak strength in comparison with those of the normal fermion 
and the antiplasmino, 
whose residues are order of unity.
It is worth mentioning that 
such smallness of the residue is actually compatible with the results in the HTL approximation:
The sum of the residues of the normal fermion and the anti-plasmino modes obtained
in the HTL approximation is unity and thus the sum rule of the spectral function of the fermion
is satisfied in the leading order.
Therefore, one could have anticipated that the residue of the ultrasoft mode 
can not be the order of unity but should be of higher order.
Equations~(\ref{eq:result-propagator}) through (\ref{eq:result-residue}) for Yukawa model
with a scalar coupling are obtained for the first time.

The pole given by Eq.~(\ref{eq:result-dispersion}) gives rise to a new peak in the spectral 
function of the fermion as
\begin{align}\label{eq:spectral}
\rho_+\pw= \frac{Z}{\pi}\im \frac{-1}{\pzero+v|\vp|+i\zeta},
\end{align}
which is depicted in the right panel of Fig.~\ref{fig:dispersion}, where $|\vp|$ is set to zero.
Since the expression of $\zeta$ for the Yukawa model is not available in the literature,
we simply adopt $\zeta=\cp^4T\ln \cp^{-1}/(2\pi)$ in the figure.

\subsection{Absence of vertex correction in Yukawa model}
\label{sec:no-vertex}
\begin{figure*}
\begin{center}
\includegraphics[width=0.8\textwidth]{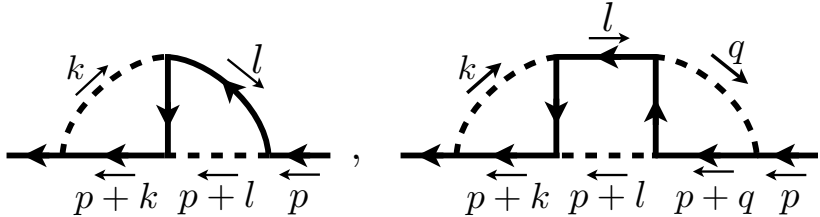}
\caption{Some of the ladder class diagrams.
The solid and dashed line are the dressed propagators of the fermion and the scalar boson, respectively.
}
\label{fig:ladder}
\end{center}
\end{figure*}

So far, we have considered the one-loop diagram. 
We need to check that the higher-order loops are suppressed by the coupling constant.
This task would not be straightforward because,  $\delta m^2\sim\cp^2 T^2$
appears in the denominator, as seen in  Eq.~(\ref{eq:one-loop-pc}),
which could make invalid the naive loop expansion. 
The possible diagrams contributing in the leading order are ladder diagrams 
shown in Fig.~\ref{fig:ladder} because
the fermion-boson pair of the propagators gives a contribution of order $1/g^2$, and the vertex gives $g^2$.
However, there is a special suppression mechanism in the present case with the scalar coupling.

For example, let us evaluate the first diagram in Fig.~\ref{fig:ladder},
at  small $p$.
The self-energy is evaluated to be
\begin{align}
\begin{split}
\simeq&\int\frac{d^4k}{(2\pi)^4}\tilde{K}(k)\int\frac{d^4l}{(2\pi)^4}\tilde{\kernel}(l)
\frac{\Slash{k}(\Slash{k}-\Slash{l})\Slash{l}}{(2k\cdot l )}
\frac{2\tilde{p}\cdot (k-l)}{\delta m^2} .
\end{split}
\end{align}
Since there are four vertices and two pairs of the propagators whose momenta are almost the same, 
the formula would apparently yield the factor, $\tilde{\kernel}(k)\tilde{\kernel}(l)\sim\cp^4\times(\delta m^2)^{-2}\sim g^0$.
One can easily verify that this order estimate would remain the same in any higher-loop diagram, 
so any ladder diagram seems as if to contribute in the same leading order as explained.
However, this is not the case for Yukawa model with the scalar coupling. 
An explicit evaluation of the numerators of the fermion propagators gives
$\Slash{k}(\Slash{k}-\Slash{l})\Slash{l}=\Slash{l}k^2-\Slash{k}l^2$, which
turns out to be of  order $\cp^2$.
This is because the internal line is almost on-shell, i.e., 
$k^2$, $l^2\sim \cp^2T^2$,
which comes from the asymptotic masses squared. 
An analysis shows that 
the same suppression occurs in the higher-order diagrams 
such as the second diagram in Fig.~\ref{fig:ladder}.
 Thus, 
the ladder diagrams giving a vertex correction 
do not contribute in the leading order in the scalar coupling, and 
hence, the one-loop diagram in Fig.~\ref{fig:oneloop}
 with the dressed propagators solely suffices 
to give the self-energy in the leading-order.

We remark that a similar suppression occurs
in the effective three-point-vertex at $p\sim gT$ \cite{scalar}.
We also note that
 this suppression mechanism is 
quite similar to that found in a supersymmetric model
for an intermediate temperature region in the sense that 
$gT\ll m$~ \cite{lebedev1}, whereas we are dealing with extremely high-$T$ case.
It should be emphasized that this suppression of the vertex correction is not the case 
in QED/QCD,
 where all the ladder diagrams contribute in the leading order 
and must be summed over~\cite{lebedev2},  as will be shown for QED in the next section.

\section{Ultrasoft Fermionic Modes in Quantum Electrodynamics}\label{sec:QED}
Next we explore whether the ultrasoft fermionic mode also exists in QED at high $T$. 
One might expect that the analysis would  be done in much the same way as in the Yukawa model.
It turns out, however,  that 
the analysis is more complicated and involved. 
It is necessary to sum up the contributions from all the ladder diagrams
even apart from the complicated helicity structure of the photon.
In this section, we successfully perform the summation of the ladder diagrams in an analytic way,
and obtain the fermion propagator that is valid in the ultrasoft region.
Then we evaluate  the pole in the ultrasoft region explicitly and
examine the properties of the ultrasoft fermionic mode in QED.
We also discuss whether the resummed vertex satisfies the Ward-Takahashi identity.

\subsection{One-loop calculation}
First, we evaluate the contribution from the one-loop diagram.
We use $g$ as a coupling constant of QED instead of the standard notation $e$
to make it clear
that the same order counting appears as that in the Yukawa model.

The dressed propagators with hard momenta read
\begin{align}
\GR(k) &\simeq \frac{-\Slash{k}}{k^2-\me^2+2i\zetae k^0} , \\
\label{eq:qed-photon-retarded-propagator}
\DR_{\mu\nu}(k)& \simeq \frac{-P^T_{\mu\nu}(k)}{k^2-\mph^2+2i\zetaph k^0},\\
\GF(k)&\simeq\left(\frac{1}{2}-n_F(k^0)\right)\Slash{k}\frac{4i\zetae k^0}{(k^2-\me^2)^2+4\zetae^2(k^0)^2},\\
\label{eq:qed-photon-s-propagator}
\DF_{\mu\nu}(k)&\simeq\left(\frac{1}{2}+n_B(k^0)\right)\frac{4i\zetaph k^0 P^T_{\mu\nu}(k)}{(k^2-\mph^2)^2+4\zetaph^2(k^0)^2} .
\end{align}
Here $m_e=gT/2$ and $m_\gamma=gT/\sqrt{6}$ are
 the asymptotic masses of electron and photon
 respectively~\cite{dispersion-numerical,Blaizot:2001nr}.
The damping rates of electron and photon are estimated as $\zeta_e\sim\cp^2T\ln \cp^{-1}$ \cite{damping-hard-electron} 
and $\zeta_\gamma\sim\cp^4T\ln g^{-1}$ \cite{damping-hard-photon}. 
Note that $\zetae$ is much larger than that in the Yukawa model, 
which is called ``anomalous damping'' \cite{damping-hard-electron}.
This large electron damping makes the damping rate of the ultrasoft mode larger than in the Yukawa model.
$P^T_{\mu\nu}(k)$ is the projection operator on the transverse direction,
\begin{equation}
P^T_{\mu\nu}(k) \equiv g_{\mu i}g_{\nu j}(\delta_{ij}-\hat{k}_i\hat{k}_j) ,
\end{equation}
with $\hat{k}^i\equiv k^i/|\vect{k}|$. 
Here we have adopted the Coulomb gauge, in which the analysis becomes simple thanks to the transversality of the photon propagator.
We note that the $u^\mu u^\nu/|\vk|^2$ term, where  $u^\mu=(1,\vect{0})$, 
has been omitted in Eqs.~(\ref{eq:qed-photon-retarded-propagator}) and (\ref{eq:qed-photon-s-propagator}) because that term vanishes after the $k^0$ integral.
The analysis in other gauge-fixing conditions are more complicated, 
and will be reported elsewhere \cite{future}.

By using these resummed propagators, 
the one-loop contribution in the ultrasoft region is evaluated as 
\begin{equation}
\label{eq:pairPropagator}
\begin{split}
\varSigma_\text {one-loop}^R(p) &=i\cp^2\int\frac{d^{4}k}{(2\pi)^{4}} \gamma^\mu \bigl[D^S_{\mu\nu}(-k)G^R(p+k)
+D^R_{\mu\nu}(-k)G^S(p+k)\bigr]\gamma^\nu\\
&\simeq2\int\frac{d^{4}k}{(2\pi)^{4}}\tilde{\kernel}(k)\frac{\Slash{k}}{1+2\tilde{p}\cdot k/\deltam^2} ,
\end{split}
\end{equation}
where $\deltam^2\equiv\mph^2-\me^2$, $\damping\equiv\zetae+\zetaph\simeq\zetae$.  
Here we have used the same notation for $\delta m^2$ and $\zeta$ as those in the Yukawa model, 
although their parametrical expressions are different from each other.
The factor two in the last line of Eq.~(\ref{eq:pairPropagator}) comes from two degrees of freedom of photon polarization.
At the one-loop order, we obtain
\begin{equation}
\begin{split}
\varSigma^R_\text{one-loop}(p) 
&= -\frac{16\pi^2}{\cp^2}\lambda^2 \big( (p^0+i\zeta)\gamma^0+v\vp\cdot \vgamma \big).
\end{split}
\end{equation}
We note that this expression has the same structure as that for the Yukawa model;
see Eq.~(\ref{eq:sigma-p1}).

\begin{figure}
\begin{center}
\includegraphics[width=0.5\textwidth]{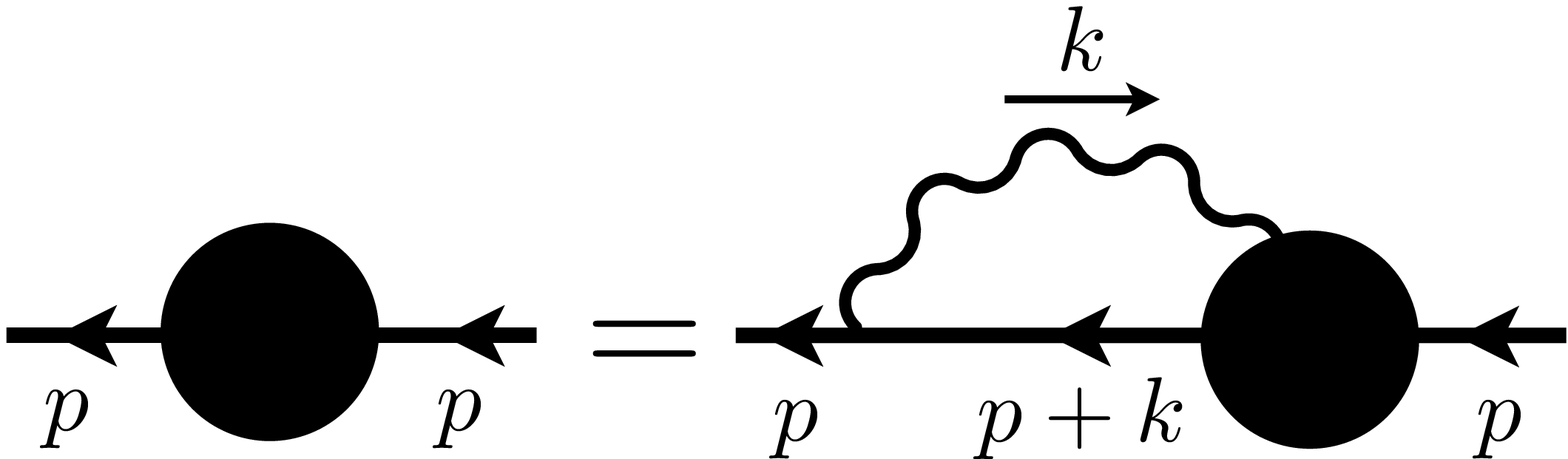}\\
\includegraphics[width=0.85\textwidth]{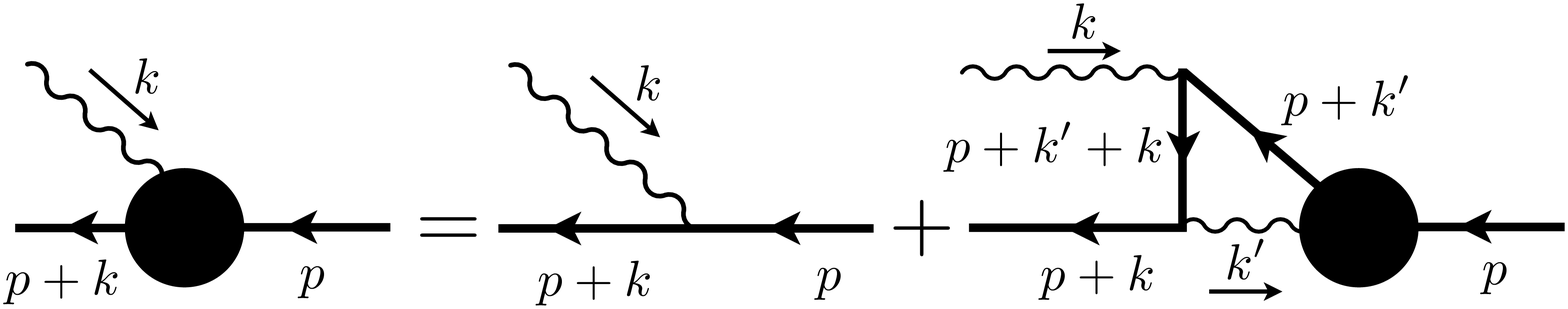}
\caption{Resummed self-energy and the self-consistent equation for the vertex function.
}
\label{fig:selfconsistentEquation}
\end{center}
\end{figure}

\subsection{Ladder summation}
As already mentioned,
 the ladder diagrams contribute to the leading-order unlike 
in the Yukawa model.
In this subsection, we 
sum up all the ladder diagrams, and 
obtain 
the analytical expressions of 
the pole position and the residue of the ultrasoft mode. 

For this purpose,
we introduce the vertex $\cp\varGamma^\mu(p,k)$
 defined through the following self-energy:
\begin{equation}
\varSigma^R(p) =
\int\frac{d^{4}k}{(2\pi)^{4}}\tilde{\kernel}(k)\frac{\gamma^{\mu}\Slash{k}P^T_{\mu\nu}(k)
\varGamma^{\nu}(p,k)}{1+2\tilde{p}\cdot k/\deltam^2}.
\label{eq:selfEnergy}
\end{equation}
Here the vertex contains the contributions from all the ladder diagrams and
satisfies the following self-consistent equation: 
\begin{align}
\varGamma^\mu(p,k)&=\gamma^{\mu}  +\int\frac{d^{4}k'}{(2\pi)^{4}}
\tilde{K}(k')\gamma^\nu\frac{\Slash{p}+\Slash{k}+\Slash{k}'}{(p+k+k')^2}
\frac{\gamma^\mu(\Slash{p}+\Slash{k}')}{1+2\tilde{p}\cdot k'/\deltam^2}\notag\\
&\qquad\qquad\times P^T_{\nu\rho}(k') \varGamma^\rho(p,k').
\label{eq:vertex}
\end{align}
Equations (\ref{eq:selfEnergy}) and (\ref{eq:vertex}) are represented diagrammatically
in Fig.~\ref{fig:selfconsistentEquation}.
We have used the same approximation as that used in the derivation of Eq.~(\ref{eq:bareSelfEnergy}) for the propagator of fermion 
and photon.
We should remark here that this summation scheme using the self-consistent equation 
was first constructed in Ref.~\cite{lebedev2}. However, 
we also note that the equation has never been solved  either analytically nor numerically.
In the following, we solve this self-consistent equation analytically for small $\tilde{p}$, 
and show that the dispersion relation does not change from that in the one-loop order 
even after incorporating all the ladder diagrams.

At small $\tilde{p}$, Eq.~(\ref{eq:vertex}) reduces to 
\begin{align}
\varGamma^\mu(p,k)
&=\gamma^\mu +\int\frac{d^4k'}{(2\pi)^4}\tilde\kernel(k')
\frac{P^T_{\nu\rho}(k')}{1+2(\tilde{p}\cdot k')/\deltam^2}
\frac{(k^\nu \gamma^\mu+k'^\mu\gamma^\nu)\Slash{k}'\varGamma^\rho(p,k')}{k\cdot k'},
\label{eq:vertex2}
\end{align}
where we have dropped the term proportional to $\Slash{k}$ because 
it only gives higher order contribution after being multiplied by $\Slash{k}$ 
in the numerator of Eq.~(\ref{eq:selfEnergy}), because
$\Slash{k}^2=k^2\sim g^2T^2$.

Let us solve the self-consistent equation (\ref{eq:vertex2}).
We expand the vertex function as
\begin{equation}
\varGamma^\mu(p,k) = \varGamma^\mu_0(k) +\delta \varGamma^\mu(p,k) ,
\end{equation}
where $\varGamma^\mu_0(k)$ is of order unity and $\delta\varGamma^\mu(p,k)$ is of order $\tilde{p}/(\cp^2T)$.

We first evaluate $\varGamma^\mu_0(k)$, which can be decomposed as follows:
\begin{equation}
\varGamma^\mu_0(k)=\gamma^\mu A(k) + k^\mu B(k)+u^\mu C(k) ,
\label{eq:varGamma}
\end{equation}
where $A$, $B$, and $C$ are $4\times 4$ matrices.
Then the self-consistent equation for $\varGamma_0^\mu(k)$ becomes
\begin{equation}
\begin{split}
\gamma^\mu A(k)+ k^\mu B(k)+u^\mu C(k) 
=\gamma^\mu
+\gamma^{\mu}\int\frac{d^4k'}{(2\pi)^4}\tilde\kernel(k')
\frac{\Slash{k}' k^\nu P^T_{\nu\rho}(k') \gamma^\rho}{k\cdot k'}A(k') ,
\end{split}
\end{equation}
where $B(k)$ and $C(k)$  in the right hand side vanish  due to transversality of the photon propagator:
$P^T_{\mu\nu}(k) k^\nu= P^T_{\mu\nu}(k) u^\nu=0$.
By assuming that $A(k)$ is a constant, the integral becomes
\begin{equation}
\int\frac{d^4k'}{(2\pi)^4}\tilde\kernel(k')
\gamma^\mu\frac{\Slash{k}' k^\nu P^T_{\nu\rho}(k') \gamma^\rho}{k\cdot k'}A = \left(-\gamma^\mu\lambda 
+\frac{2k^\mu}{k^0}\gamma^0\lambda\right)A,
\label{eq:integral}
\end{equation}
where we have dropped $\Slash{k}$ term and imposed $k^2=0$ as before.
Then, we find
\begin{align}
A = \frac{1}{1+\lambda}\vect{1},
\quad B(k) = \frac{1}{k^0}\frac{2\lambda}{1+\lambda}\gamma^0,
\quad C(k)=0,
\label{eq:varGammaABC}
\end{align}
where $\vect{1}$ is the unit $4\times 4$ matrix. 

Next we evaluate $\delta\varGamma(p,k)$.
From Eq.~(\ref{eq:vertex2}), expanding $\varGamma^\mu(p,k)$ in terms of $2\tilde{p}\cdot k'/\delta m^2$, we find the self-consistent equation for $\delta\varGamma(p,k)$ as
\begin{equation}
\begin{split}
\delta\varGamma^\mu(p,k)=& \int\frac{d^4k'}{(2\pi)^4}
\tilde\kernel(k')P^T_{\nu\rho}(k')
\frac{(k^\nu \gamma^\mu+k'^\mu\gamma^\nu)\Slash{k}'}{k\cdot k'} \\
&\qquad\qquad\times\left[-A\gamma^\rho\frac{2\tilde{p}\cdot k'}{\deltam^2}+\delta\varGamma^\rho(p,k')\right].
\end{split}
\label{eq:deltaVarGamma}
\end{equation}
It is not easy to analytically solve Eq.~(\ref{eq:deltaVarGamma}) directly.
So we instead calculate the following function:
\begin{equation}
\label{eq:deltaPi}
\delta\varPi(p)=
\int\frac{d^4k}{(2\pi)^4}\tilde\kernel(k)P^T_{\mu\nu}(k)  \gamma^\mu \Slash{k}\delta \varGamma^\nu(p,k) .
\end{equation}
Then Eq.~(\ref{eq:deltaVarGamma}) leads to the following closed equation,
\begin{equation}
\begin{split}
\delta\varPi(p)=&
-4 A\int\frac{d^4k}{(2\pi)^4}\tilde\kernel(k)\int\frac{d^4k'}{(2\pi)^4}\tilde\kernel(k')P^T_{\mu\nu}(k)  \gamma^\mu \Slash{k}
\frac{k'^\nu\Slash{k}'}{k\cdot k'} 
\frac{\tilde{p}\cdot k'}{\deltam^2}\\
&\quad +\int\frac{d^4k'}{(2\pi)^4}
\left[ 
\int\frac{d^4k}{(2\pi)^4}\tilde\kernel(k)P^T_{\mu\nu}(k)  \gamma^\mu \Slash{k}\frac{k'^\nu}{k\cdot k'}\right]\\
&\qquad\qquad\qquad\times\tilde\kernel(k') P^T_{\rho\lambda}(k')\gamma^\rho\Slash{k}'\delta\varGamma^\lambda(p,k') \\
=&-\lambda A\varSigma_\text{one-loop}(p) -\lambda\delta\varPi(p).
\end{split}
\end{equation}
Here we have used Eqs.~(\ref{eq:integral}) and (\ref{eq:deltaPi}) in the second line.
The solution to this equation is readily found to be
$\delta\varPi(p)=-\lambda A^2\varSigma_\text{one-loop}(p)$, where
Eq.~(\ref{eq:varGammaABC}) is used for $A$.
Then, the self-energy is evaluated as 
\begin{equation}
\label{eq:selfenergy-1st}
\begin{split}
\varSigma^R(p)&= \int\frac{d^4k}{(2\pi)^4}\tilde\kernel(k)P^T_{\nu\rho}(k)  \gamma^\nu \Slash{k}
\left(-\frac{2\tilde{p}\cdot k}{\deltam^2}\varGamma^\rho_0(k)+\delta \varGamma^\rho(p,k)\right)\\
&=A \varSigma_\text{one-loop}(p)
+\delta\varPi(p) \\
&=-\frac{1}{Z}(\gamma^0(p^0+i\zeta)+v\vp\cdot \vgamma),
\end{split}
\end{equation}
where the residue is
\begin{align}
Z&=\frac{\cp^2}{16\pi^2 \lambda^2}(1+\lambda)^2.
\end{align}
The pole of the ultrasoft fermionic mode has  the velocity $v=1/3$, 
damping rate $\damping$, and the residue $Z$.
The dispersion of the mode is the same as that in the one-loop level 
whereas the residue is changed.
This is our main result for QED. 

\subsection{Ward-Takahashi identity}
In this subsection, 
we examine whether the summation scheme, Eqs.~(\ref{eq:selfEnergy}) and (\ref{eq:vertex}), 
is consistent with the $U(1)$ gauge symmetry. 
Concretely, we check that our resummed vertex function and self-energy satisfy the 
Ward-Takahashi (WT) identity in the leading order of the coupling constant.

The WT identity reads
\begin{equation}
\begin{split}
k^\mu \varGamma_{ \mu}(p,k)
&=\Slash{p}+\Slash{k}-\varSigma^R(p+k) -\Slash{p}+\varSigma^R(p).
\label{eq:WTidentity}
\end{split}
\end{equation}
Since $\varGamma_{ \mu}(p,k)$ contains two separated scales $k\sim T$ and $p\lesssim \cp^2T$, we need to treat them carefully.
For the hard part, $\varSigma^R(p+k)\simeq \varSigma^R(k)$ is of order $\cp^2T$, which is negligible compared to $\Slash{k}$.
In addition, the momentum dependent part $\varSigma^R(p+k)-\varSigma^R(k)\sim g^2 p$ is also negligible compared to $\varSigma^R(p)$.
Therefore  the WT identity reduces to  
\begin{equation}
k_\mu \varGamma^{ \mu}(p,k)=\Slash{k}+\varSigma^R(p).
\label{eq:mod-WT}
\end{equation}

On the other hand, multiplying Eq.~(\ref{eq:vertex}) by $k_\mu$, we have
\begin{equation}
\begin{split}
k_\mu\varGamma^\mu(p,k)
&=(\Slash{p}+\Slash{k})-\Slash{p}  + \int\frac{d^{4}k'}{(2\pi)^{4}}
\tilde K(k')\gamma^\nu\frac{ \Slash{p}+\Slash{k}+\Slash{k}'}{(p+k+k')^2} \\
&\quad\qquad\qquad\times\frac{((\Slash{p}+\Slash{k}+\Slash{k}')-\Slash{p}-\Slash{k}')(\Slash{p}+\Slash{k}')}{1+2\tilde{p}\cdot k'/\deltam^2}
P^T_{\nu\rho}(k') \varGamma^\rho(p,k')\\
&=\Slash{k} + \int\frac{d^{4}k'}{(2\pi)^{4}}
\tilde K(k')\gamma^\nu\left(\Slash{p}+\Slash{k}'-\frac{ \Slash{p}+\Slash{k}+\Slash{k}'}{2k\cdot k'}(p+k')^2\right) \\
&\quad\qquad\qquad\times\frac{1}{1+2\tilde{p}\cdot k'/\deltam^2}
P^T_{\nu\rho}(k') \varGamma^\rho(p,k')\\
&\simeq\Slash{k}+\varSigma^R(p),
\end{split}
\end{equation}
where we have dropped the terms of order $g^2T$, and used Eq.~(\ref{eq:selfEnergy}) in the last line.
This expression coincides with Eq.~(\ref{eq:mod-WT}),
so our self-consistent equation satisfies the WT identity.
We note that this proof was made without using the expansion in terms of $\tilde{p}/ \cp^2T$.
For this reason, Eq.~(\ref{eq:mod-WT}) is generally valid for $p\lesssim \cp^2T$,
 as well as for $\tilde{p}\ll \cp^2T$.

Next, we check whether the explicit solution Eq.~(\ref{eq:varGammaABC}) of the self-consistent 
equation (\ref{eq:vertex}) at zeroth order in $\tilde{p}/(\cp^2T)$ 
 satisfies the WT identity to see the consistency with the gauge symmetry 
of the following two conditions adopted to obtain Eq.~(\ref{eq:varGammaABC}):
One is that terms proportional to $\Slash{k}$ in $\varGamma(p,k)$ is dropped, 
since they are negligible in the self-energy due to $\Slash{k}\Slash{k}=k^2\sim g^2T^2$.
The other is that we imposed the on-shell condition $k^2=0$, 
because the internal photon in the self-energy is almost on-shell.
Using the same conditions, we expect that the vertex function satisfies the WT identity 
in the leading order of the coupling constant.
In fact, we have for the zeroth order  in $\tilde{p}$ 
\begin{equation}
\begin{split}
k_\mu \varGamma_0^{\mu}(k)& \simeq \Slash{k}A+k^2B(k)\simeq0 .
\end{split}
\end{equation}
Here  we have dropped $\Slash{k}$ and $k^2$
in the last equality.

Finally, we check that the equation determining the vertex function Eq.~(\ref{eq:deltaVarGamma}),
which is first order in $\tilde{p}$, satisfies the WT identity.
By multiplying this equation by $k_\mu$, we obtain
\begin{equation}
\begin{split}
k_\mu\delta\varGamma^\mu(p,k)&= \int\frac{d^4k'}{(2\pi)^4}
\tilde\kernel(k')P^T_{\nu\rho}(k')\gamma^\nu\Slash{k}'
\left[-A\gamma^\rho\frac{2\tilde{p}\cdot k'}{\deltam^2}+\delta\varGamma^\rho(p,k')\right]\\
&=\varSigma^R(p),
\end{split}
\end{equation}
where  we have dropped $\Slash{k}$ as in the previous equation and used Eq.~(\ref{eq:selfenergy-1st}).
Therefore, our analytic solution of the self-consistent equation satisfies the WT identity in the leading order of the coupling constant.

We note that without the summation of the ladder diagrams, the WT identity is not satisfied when the external momentum of fermion is ultrasoft.
By contrast, the ladder summation was unnecessary in the Yukawa model, in which the gauge symmetry is absent.

\section{Discussion on Origin of Ultrasoft Fermionic Mode}
\label{sec:discussion}

Let us discuss the physical origin of the ultrasoft mode for
clarifying its possible universality or robustness at high temperature.

As mentioned in Introduction, an ultrasoft fermionic mode with a vanishing mass can appear
as a phonino associated with the spontaneously broken supersymmetry 
at finite $T$~\cite{phonino,lebedev1,lebedev2}, below $T_c$:
In the massive Wess-Zumino model, 
 the supersymmetric cancellation of the fermion mass
is essential to make the phonino excitation~\cite{lebedev1}.
In fact, the massless fermionic mode is realized  as a consequence of an exact cancellation of  
the self-energy with  the finite bare  mass at the vanishing external momentum due to
the spontaneous breaking of supersymmetry.
This is quite in contrast to the case of the Yukawa model dealt in the present work,
where the ultrasoft fermionic modes do {\it not} appear 
when the fermion mass is large.
This means that the ultrasoft fermionic mode like the phonino can appear 
even in a nonsupersymmetric model like the Yukawa model, 
and there the masslessness of the fermion is an essential ingredient 
in realizing it.
Nevertheless it is interesting that such an exotic fermionic mode can  appear
both in supersymmetric and nonsupersymmetric models with quite different mechanisms at high temperature.

For getting further intuition into the possible mechanism for realizing the ultrasoft fermionic
mode in the Yukawa model\footnote{The mechanism of realizing the ultrasoft fermionic mode in QED is expected to the same as that in the Yukawa model because  the multi-loop diagrams in QED affect only the expression of the residue.}, 
we  note that
Kitazawa {\it{et al.}}  argued that the level repulsion due to the Landau damping
 causes the three-peak structure in Ref.~\cite{3peak},
 though in the case where a massless fermion is coupled with a {\it massive} boson and
thus the fermion-boson mass difference squared $\delta m^2$ is nonzero.
Indeed 
a finite $\delta m^2$ plays an important role in 
the appearance of the ultrasoft mode even when the boson is massless
because it ensures the smoothness of the self-energy at the origin:
As discussed in Introduction, 
the real part of the self-energy vanishes at the origin from the symmetry 
for the massless fermion. 
The residue of the pole is proportional to $(\delta m^2)^2$ and vanishes if 
$\mf=\mb$ in the Yukawa model or $\me=\mph$ in QED.
In the present work,  the difference in the masses squared 
becomes finite because the effects beyond the HTL approximation is taken into account
and found to be $\delta m^2\sim g^2 T^2$.
The mass difference in turn causes the smallness of the imaginary part:
The imaginary part of the self-energy can originate from 
the boson emission, the Landau damping, and the imaginary part 
of the dressed propagators obtained beyond the HTL approximation. 
The former two contributions are found to be zero at the origin \cite{3peak,shk} due to
 the nonzero mass difference or kinematics at the one-loop order.
Therefore, the leading contribution of the imaginary part is solely given by the damping rates
of the hard particles. Thus the ultrasoft fermionic mode gets to exist without supersymmetry
at high temperature.

\section{Summary and Concluding Remarks}
\label{sec:summary}

\begin{table}[t]
\caption{The expressions of the dispersion relation, the damping rate,  and the residue of the ultrasoft mode in the Yukawa model and QED.
}
\begin{center}
\begin{tabular}{l|c|c}
\hline
 & Yukawa model & QED \\ \hline \hline
dispersion relation & $\pm|\vp|/3$ & $\pm|\vp|/3$  \\
damping rate & $\zeta_f+\zeta_b \sim \cp^4T\ln \cp^{-1}$ & $\zetae\sim\cp^2T\ln \cp^{-1}$ \\
residue & $\cp^2/(72\pi^2) $ & $\cp^2/(144\pi^2)$ \\
\hline
\end{tabular} 
\end{center}
\label{tab:expression}
\end{table}

We have investigated the spectral properties of massless fermion with the ultrasoft momentum ($p \lesssim\cp^2T$) 
in Yukawa model and QED in the Coulomb gauge at high temperature.
We have first indicated that a resummed perturbation theory \cite{lebedev1,lebedev2}
is needed  beyond the conventional HTL approximation to get a sensible 
spectral properties in the ultrasoft region:
For a consistent calculation in the relevant order of the coupling,
we have shown that the use of the dressed propagators with the asymptotic
masses and the decay widths for the both fermion and boson
is necessary, and the vertex corrections
due to the summation of ladder diagrams is to contribute in the leading order.
In QED, they turn out to be of leading order and are non-negligible, while it is not the case in the Yukawa model.
Consequently, the WT identity is satisfied in QED.

Thereby we have established  that 
the resulting fermion propagator develops a pole at ultrasoft momentum, 
which was first suggested in \cite{lebedev1} for supersymmetric models in different temperature
region from ours, and in \cite{lebedev2} for gauge theory in the same temperature region as ours:
Its pole position, width, and the residue have been obtained for the first time
in the present work as a function of the asymptotic masses and damping rates; 
they are summarized in Table.~\ref{tab:expression}.
We note that we have obtained the pole of the propagator by expanding the propagator 
around $\pzero=-i\zeta$, $|\vp|=0$ instead of $\pzero=|\vp|=0$  in QED, where
 the damping rate of the hard electron is anomalously large~\cite{damping-hard-electron} unlike the Yukawa model. 
If one expands the propagator around $\pzero=|\vp|=0$, 
it is assumed that $p\ll \delta m^2/T$.  The pole that we have found has the imaginary 
part $\im \pzero=\zeta \sim g^2T$, so the expansion would break down because  $pT/\delta m^2\sim 1$ at the pole position.
We also remark that although the present analysis is given for a massless fermion, 
 the ultrasoft fermionic excitation should exist
 even for a massive fermion, if the bare mass  is smaller
than $\cp^2T$, which is the smallest scale in our analysis.

We should also emphasize that 
 the existence of a peak in the fermion spectral function at the origin 
implies that the fermion spectral function has a three-peak structure,
although the central peak has only a small strength.
The development of such a three-peak structure at intermediate
temperature was suggested 
in the case where the boson is massive irrespective of the type 
of the boson~\cite{3peak,mitsutani,shk}, as mentioned in Introduction.
 Thus, we see that the three-peak structure of the fermion spectral
function is a robust phenomenon to be seen 
at both of intermediate and high $T$, at least in the 
Yukawa model.
In QED, further analysis is needed to establish the existence of the peak in the spectral function at high $T$.
In general, the propagator can be expanded by the partial fractions~\cite{newton}, and
several poles near the real axis contributes to the spectral function, although infinite number of poles generally exist in complex energy plane.
The pole that we have found in this paper is one of them.
It is, therefore, needed to examine whether the contributions from other poles are small\footnote{
In Yukawa model, smallness of damping rate $\zeta$ justifies the existence of the peak of spectral function at small $g$.}.

Although the present analysis is restricted to the Yukawa model with a scalar coupling 
and QED, 
a similar analysis can be performed in QCD~\cite{lebedev2,future}.
We note, however, that the self-coupling between the gluons 
leads to additional ladder diagrams that should be also summed up~\cite{lebedev2}. 
The detailed analysis will be reported elsewhere~\cite{future}.

It is known that the HTL approximation is equivalent to the Vlasov equation \cite{blaizot-HTL}. For the bosonic channel in ultrasoft or softer momentum region, the ladder summation is required and  corresponding kinetic theory is the Boltzmann equation \cite{ultrasoft-am}.
It is interesting to develop a kinetic equation for fermionic channel in the ultrasoft momentum region, which will be discussed in our future work~\cite{future}.

\section*{ACKNOWLEDGMENTS}
Y. H. thanks Robert D. Pisarski for fruitful discussions and comments.
This work was supported by the Grant-in-Aid for the Global COE Program 
``The Next Generation of Physics, Spun from Universality and Emergence'' from the Ministry of Education, 
Culture, Sports, Science and Technology (MEXT) of Japan and by a Grant-in-Aid 
for Scientific Research by the Ministry of Education, Culture, Sports, Science and Technology (MEXT) 
of Japan (No. 20540265 and No.2334006).


\end{document}